\documentclass[aps,prb,twocolumn,superscriptaddress,10pt,showpacs]{revtex4}

\usepackage{amsmath}
\usepackage{epsfig}
\usepackage{graphics}

\usepackage{float}

\newcommand{\br}[1]{\breve{#1}}

\newcommand{\arcosh}{{\rm arcosh}}
\newcommand{\artanh}{{\rm artanh}}

\newcommand{\abs}[1]{{\left|#1\right|}}
\renewcommand{\Im}{{\rm Im}}
\renewcommand{\Re}{{\rm Re}}

\def\ba{\begin{array}}
\def\ea{\end{array}}
\def\l{\left}
\def\r{\right}

\begin{document}


\title{Supercurrent-carrying density of states in diffusive mesoscopic
Josephson weak links}


\author{Tero T. Heikkil\"a}
\email[Author to whom correspondence should be addressed. Electronic
address:]{Tero.T.Heikkila@hut.fi}
\altaffiliation{Present address: Low Temperature Laboratory, P.O. Box
2200, FIN-02015 HUT, Finland}

\affiliation{Materials Physics Laboratory, Helsinki University of
Technology, P.O. Box 2200, FIN-02015 HUT, Finland}
\affiliation{Institut f\"ur Theoretische Festk\"orperphysik, Universit\"at Karlsruhe, D-76128 Karlsruhe, Germany}

\author{Jani S\"arkk\"a}
\affiliation{Materials Physics Laboratory, Helsinki University of
Technology, P.O. Box 2200, FIN-02015 HUT, Finland}

\author{Frank K. Wilhelm}
\affiliation{Sektion Physik and CeNS, Ludwig-Maximilians-Universit\"at, 
Theresienstr. 37, D-80333 M\"unchen, Germany}


\date{\today}

\begin{abstract}
Recent experiments have demonstrated the nonequilibrium
control of the supercurrent through diffusive phase-coherent
normal-metal weak links. The experimental results have been accurately
described by the quasiclassical Green's function technique in the
Keldysh formalism. Taking into account the geometry of the structure,
different energy scales and the nonidealities at the interfaces allows us
to obtain a quantitative agreement between the theory and the
experimental results in both the amplitude and the phase dependence of 
the supercurrent, with no or very few fitting parameters. Here we discuss
the most important factors involved with such comparisons: the ratio
between the superconducting order parameter and the Thouless energy of 
the junction, the effect of additional wires on the weak link, and the 
effects due to imperfections, most notably due to the nonideal
interfaces. 
\end{abstract}


\maketitle

\section{Introduction}
\label{Introduction}
Many quantum phenomena in many-body systems are based on probing the spectrum of 
states corresponding to the desired observable, the states being filled
according to an appropriate distribution function. A similar viewpoint can be 
taken also on the
Josephson effect: supercurrent is carried by states in the weak link
and their occupation is determined by a distribution function
antisymmetric between the electron and hole spaces. This aspect is 
directly reflected in the mathematical structure of the supercurrent 
formula derived from the Keldysh Green's-functions
method.\cite{wilhelm,yip,volkov}  
Such an approach has been taken in some recent
experiments\cite{morpurgo,baselmans,huang,schaepers,baselmanssquid,kutchinsky} 
controlling the Josephson effect in phase-coherent normal-metal wires
through the control of the distribution function by an injection of
normal quasiparticle 
current. One of the most remarkable results of these experiments is
the inversion of the sign of the supercurrent for a given phase
difference across the weak link when the junctions turn into a
$\pi$ state. 

Quantitative fit to the experimentally obtained results has been very 
successful for the equilibrium supercurrent\cite{Pascal} using the
equilibrium quasiclassical theory. In the nonequilibrium case, detailed 
knowledge of the relaxation mechanisms controlling the shape of the
interactions, but also the precise spectrum of supercurrent-carrying
states, is required.\cite{huang,baselmanssquid}
Previously,\cite{wilhelm}
for the calculation of this spectrum, one has
assumed a two-probe setup with some idealized conditions on the length
scales and on the nature of the interfaces. In this paper, we systematically
investigate
the spectrum of this current-carrying density of states, or spectral
supercurrent, show how it is calculated, and how it depends on the
length of the weak link, presence of additional terminals, or on the
nonidealities in the interfaces between the normal-metal weak link and
the superconductors. We also discuss the current-phase relation of
such a system: at low temperatures, it can be far from sinusoidal, and
at certain conditions, its period can even be
halved.\cite{baselmansphase} We focus on the diffusive limit 
where the dimensions $d$ of the weak link are much greater than the
elastic mean free path $l$. This is the typical limit for most
normal-metal weak links. The corresponding ballistic limit $d \ll l$
has been extensively described in the
literature\cite{kulik,ishii,bardeenjohnson,gunsenheimer,beenakkervanhouten,buttikerklapwijk,vanwees91}
in terms of Andreev bound states (ABS). We show qualitatively a
connection between the discrete ABS and the continuous diffusive-limit
spectral supercurrent.

This paper is organized as follows. After this introduction, Sec.\ II
introduces to the theoretical formalism which is based on the
real-time Usadel equation for the quasiclassical Green's function in
the diffusive limit.\cite{usadel,review} In the case of nonideal interfaces
or in multiterminal  geometries, the boundary 
conditions to these functions are also essential. Understanding the
results of the following sections does not require a detailed reading
of this part but it is enough to grasp the idea of the relation of the
spectral supercurrent and the observable one. In Sec.\ III we look
how the spectral supercurrent depends on the length of the weak link
compared to the superconducting coherence length and separate two
extreme cases. In the limit of a short junction where the coherence
length is much longer than the weak link, one obtains an analytical
solution for the spectral supercurrent without further
approximations. The current-phase relation in diffusive normal-metal
weak links is considered in Sec.\ IV. We show how, especially at low
temperatures, higher harmonics appear in addition to the usual
sinusoidal phase dependence and indicate how the period can be halved
in a nonequilibrium situation. Section V considers the effect of
additional normal-metal terminals on the current-carrying density of
states, and in Sec.\ VI, we discuss how nonidealities in the normal
metal --- superconductor (NS) interface change its shape. Finally, in 
Sec.\ VII, we summarize the main results.

\begin{figure}[h]
\includegraphics[width=\columnwidth]{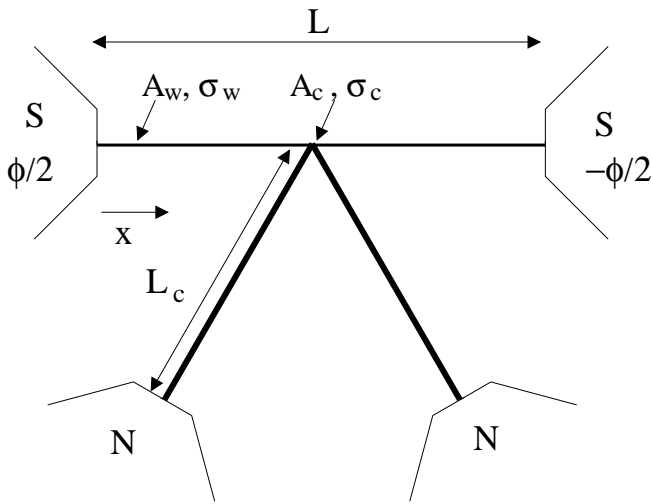}
\caption{Multiterminal SNS Josephson junction. The weak link consists
of a phase coherent normal-metal wire of length $L$, cross section
$A_w$, and normal-state conductivity $\sigma_w$, yielding a normal-state
resistance $R_N=L/\sigma_wA_w$. Additional
normal-metal wires of length $L_c$, total cross section $A_c$, and
normal-state conductivity $\sigma_c$, i.e., with resistance
$R_c=L_c/\sigma_wA_w$, called the control wires, are connected to the
center of the weak link, and from their other end, to normal
reservoirs.} 
\label{fig:structure}
\end{figure}

To be specific, we consider the structure shown in Fig.\
\ref{fig:structure}. The main wire with length $L$ and cross section
$A_w$ between the superconductors forms the weak link whereas the
additional wires with length $L_c$ and integrated cross section $A_c$
are used for the control of the distribution functions and therefore
referred to as the control wires. We assume that the superconducting
and normal reservoirs are much larger than the weak link and the
control wire, such that the Green's functions describing them take
their bulk values very close to the interfaces. Furthermore, we assume 
that the width of the control wires is much smaller than the length
$L$ of the weak links. This allows us to consider the wires as 
quasi-one-dimensional structures by assuming translational invariance
in the transverse directions. 

\section{Theoretical background}
\label{theory}
Circuits composed of normal and superconducting metals
 in the diffusive limit (dimensions larger than the 
elastic mean free path $l$) are effectively described in terms of the
quasiclassical Green's functions $\br{G}$ satisfying the Usadel
equations\cite{usadel,review}
\begin{equation}
D\nabla (\br{G}\nabla
\br{G})=\left[-i(E+i\Gamma)\br{\tau_3}+\br{\Delta},\br{G}\right], 
\label{eq:usadeldiff}
\end{equation}
where $D=\frac{1}{3}v_F l$ is the diffusion constant, $E$ is energy relative
to the chemical potential of the superconductors (which is assumed to
be the same for all S terminals),
$\Gamma$ describes a small inelastic scattering rate, and $\br{\Delta}$
the superconducting pair potential (we set $\hbar=1$ 
throughout). Since we aim to describe nonequilibrium effects, we adopt 
the Keldysh real-time formalism\cite{rammersmith} and hence
\begin{equation}
\br{G}=\l(\ba{cc} \hat{G}^R & \hat{G}^K\\0 &
\hat{G}^A\ea\r), \quad 
\br{\Delta}=\l(\ba{cc} \hat{\Delta} & 0 \\ 0 &
\hat{\Delta}\ea\r),\quad
\br{\tau}_3=\l(\ba{cc} \hat{\tau}_3 & 0 \\ 0 & \hat{\tau}_3
\ea\r).
\end{equation}
All of the submatrices denoted by a hat ($\hat{G}^R$, etc.) are
$2\times 2$ matrices in Nambu particle-hole space, in particular, 
$\hat{\tau}_3$ is the third Pauli matrix and $\hat{\Delta}$ has the
form 
\begin{equation}
\hat{\Delta} = \l(\ba{cc} 0 & \Delta(x) \\ \Delta^*(x) &
0\ea\r).
\end{equation}
The pair potential $\Delta(x)$ can in principle be obtained from a 
self-consistency relation.\cite{review,solsferrer} However,
since we consider only superconducting reservoirs much wider than the
weak link, we adopt the usual step-function form for $\Delta(x)$
(finite constant in the superconductors, zero in the normal-metal
wires).  

In addition to Eq.\ (\ref{eq:usadeldiff}), Usadel Green's function
satisfies a normalization condition $\br{G}^2=\check{1}$. Therefore it can 
be parametrized with four scalar parameters as follows.\cite{review}
The Keldysh Green's function $\hat{G}^K$ 
describing the occupation numbers of different quantum states, i.e.,
the (non)equilibrium state of the system can be expressed with two
real distribution functions $f_L$ and $f_T$ as $\hat{G}^K=\hat{G}^R
(f_L+f_T\hat{\tau}_3) - (f_L+f_T\hat{\tau}_3)\hat{G}^A$ whereas the
retarded and advanced Green's functions, $\hat{G}^R$ and $\hat{G}^A$,
describing the spectral properties which do not directly depend on the
distribution functions are
\begin{equation}
\hat{G}^R = \l(\ba{cc} \cosh(\theta) & \sinh(\theta)\exp(i\chi)\\
-\sinh(\theta)\exp(-i\chi) & -\cosh(\theta)\ea\r)
\end{equation}
and $\hat{G}^A=-\hat{\tau}_3 (\hat{G}^R)^\dagger \hat{\tau}_3$. Here
$\theta(x;E)$ and $\chi(x;E)$ are in general complex scalar functions.

In what follows, we describe a quasi-one-dimensional situation, where
the functions are assumed to vary only in one dimension 
$x$. Expressing the coordinate $x$ in terms of the separation $L$ of
the superconductors between which the supercurrent flows, $x\equiv
x'L$, the spectral equations for $\hat{G}^{R(A)}$ read in a normal
metal ($\Delta=0$) 
\begin{align}
\partial_{x'}^2 \theta &= -2i(E'+i\Gamma') \sinh(\theta)
+ \frac{1}{2} (\partial_{x'} \chi)^2 \sinh(2\theta)\label{eq:usadela},\\
j_E &\equiv -\sinh^2(\theta) \partial_{x'} \chi, \quad \partial_{x'} j_E = 0.
\end{align}
Here, the prime over the (dimensionless) quantities denotes the fact
that the energies are expressed in the units of the Thouless energy
$E_T=D/L^2$ corresponding to the length $L$. Below, we tacitly assume
all lengths and energies expressed in these natural units even if not
marked by a prime. The kinetic equations satisfied by the distribution
functions $f_L$ and $f_T$ are described, e.g., in Ref.\
\onlinecite{review}, where the part of the distribution function which is
symmetric 
about the chemical potential of the superconductors
corresponds to $f_T$ and the antisymmetric part to $f_L$. These two 
components acquire different space and energy dependent diffusion
coefficients due to the superconducting proximity effect.  

If the interfaces to the reservoirs are ideal metallic, the 
parameters are continuous at the boundaries to the reservoirs and
can be identified with the bulk values,
$\theta_S={\rm artanh}(\Delta/E)$ and $\theta_N=0$ in the superconducting
and normal-metal reservoirs, respectively. In general, e.g., if a 
supercurrent is driven through the system, 
there can be a phase difference, which we choose to be applied symmetrically
between the superconductors, such that in the
left superconductor $\chi=\phi/2$ and in the right
$\chi=-\phi/2$. Below, if not mentioned otherwise, we choose
$\phi=\pi/2$, which typically yields a supercurrent close to the
critical current of the junction.

Nonideal interfaces with reduced transmissivities are
not directly described by the Usadel equation, because 
they are of microscopic, atomic-scale thickness. 
They can, however, be taken into account using
boundary conditions derived by
Zaitsev\cite{zaitsev} for Eilenberger Green's functions (valid
independent of the mean free path) and later simplified in the
diffusive limit by Kuprianov and Lukichev 
for a tunneling case\cite{kuprianovlukichev} and Nazarov for a
general interface,\cite{nazarovbc} described by a scattering
matrix. For an interface characterized by the transmission eigenvalues
$T_n$, the Green's functions $\br{G}_1$ on the right-hand side and
$\br{G}_2$ on the left-hand side of the interface
satisfy\cite{nazarovbc,belzigspin} 
\begin{equation}
\begin{split}
\sigma_N^1 A_1 \br{G}_1 \partial_x \br{G}_1 &= \sigma_N^2 A_2 \br{G}_2 \partial_x
\br{G}_2  \\
&=\frac{2e^2}{\pi} \sum_n \frac{T_n [\br{G}_1,\br{G}_2]}{4+T_n(\{\br{G}_1,\br{G}_2\}-2)},
\end{split}
\label{eq:nazarovbcforgf}
\end{equation}
evaluated at the position of the interface. In most cases, the
individual transmission eigenvalues are not known, but since typical
interfaces contain a huge number of channels, it is enough to
integrate over the probability distribution of the eigenvalues to
obtain the desired boundary condition.

In the case of a
tunneling interface (where all the transmission eigenvalues of the
interface are small) the boundary conditions between the parametrized
functions in wires 1 and 2 reduce to\cite{review,kuprianovlukichev} 
\begin{align}
\partial_x \theta_1 = [\sinh(\theta_1)& \cosh(\theta_2)
- \sinh(\theta_2)\cosh(\theta_1)\cos(\Delta\chi)]/r_b,\\
\sinh^2(\theta_1)\partial_x \chi_1 &= \sinh(\theta_1) \sinh(\theta_2)
\sin(\Delta \chi)/r_b.
\label{eq:tunnelconditions}
\end{align}
Here, $\Delta\chi \equiv \chi_1-\chi_2$ and $\theta_{1(2)}
\equiv\theta(x_b^{+(-)})$ and $\chi_{1(2)} \equiv \chi(x_b^{+(-)})$
are the parameters $\theta$ and $\chi$ at the interface, $x=x_b$, but
on the side of the wire 1 (2). The nonideality of the interface is
characterized by the ratio of its resistance $R_I$ and of the
weak-link resistance $R_N$, $r_b \equiv R_I/R_N$ and the derivatives
point towards the wire  1. In the case of a dirty interface, where 
the boundary condition is evaluated using the distribution function of
the transmission eigenvalues corresponding to an interface with a
random array of scatterers in a 2D layer,\cite{schep} we get
\begin{align}
\partial_x \theta_1 = &\frac{\sqrt{2}[\sinh(\theta_1) \cosh(\theta_2)
- \sinh(\theta_2) \cosh(\theta_1) \cos(\Delta \chi)]}{r_b {\cal D}},\\
\sinh^2(\theta_1)& \partial_x \chi_1 = \frac{\sqrt{2} \sinh(\theta_2)
\sinh(\theta_1) \sin(\Delta \chi)}{r_b {\cal D}}.
\end{align}
Here we denoted the denominator ${\cal D}\equiv \sqrt{1+\cosh(\theta_2)
\cosh(\theta_1) - \sinh(\theta_2) \sinh(\theta_1) \cos(\Delta
\chi)}$. This denominator reflects the contribution of open 
conduction channels which are not present in
Eq.~(\ref{eq:tunnelconditions}).  

Note that both types of boundary conditions indicate a form of a
conservation of a spectral current over the interface, the second
equation being the conservation of the spectral supercurrent $j_E$. 

In geometries with more than two terminals, we assume that
narrow quasi-one-dimensional wires connect to each other at some point 
of the structure. Therefore we need to impose appropriate matching
conditions.\cite{zaitsevmc,nazarovbc,gwz,review} In this case, they
are the continuity of the 
functions $\theta$ and $\chi$ and the conservation of the
spectral currents. Assuming that the derivatives in the $N$
wires $i=1,... N$ with cross sections $A_i$ and normal-state
conductivities $\sigma_N^i$ point towards the crossing point at $x_c$,
we get
\begin{align}
\theta_i (x_c) = \theta_j (x_c) \quad &\forall i,j = 1,... N,\\
\chi_i (x_c) = \chi_j (x_c) \quad &\forall i,j = 1,... N,\\ 
\sum_{i=1}^N A_i \sigma_N^i \partial_x \theta_i(x_c) &= 0,\\
\sum_{i=1}^N A_i \sigma_N^i \partial_x \chi_i(x_c) &= 0 \label{eq:specsupcons}.
\end{align}
In the last condition we used the continuity of the parameters
$\theta$ across the crossing point.

Below, we assume the system depicted in Fig.\ \ref{fig:structure}: two
superconductors connected by ``horizontal'' mesoscopic normal wires to
which we connect normal reservoirs by the ``vertical'' mesoscopic
normal wires (labeling of the wires as in Fig.\
\ref{fig:structure}). When considering the supercurrent between the two
superconductors, for the spectral equations it is enough to treat any
number of ``vertical'' wires by a single wire for which the product of 
$\sigma_N A$ is simply the sum of these products in the individual
wires. In the case that the dependence on the length $L_c$ of these
wires becomes important, the smallest of them characterizes the
situation the best. In this case, since there can be no supercurrent
flowing to the normal reservoirs, Eq.\ (\ref{eq:specsupcons}) reduces
to $j_E^1 = - j_E^2$. Furthermore, for simplicity, we
assume the system left-right symmetric, such that the part of the weak
link in the left-hand side of the cross is similar to that in the
right-hand side.

Finally, the observable supercurrent is obtained from the solutions to 
the spectral and kinetic equations by
\begin{equation}
I_S=\frac{E_T}{2eR_N} \int_{-\infty}^\infty dE'
\Im\left[j_E(E')\right] f_L(E'). 
\label{eq:observableis}
\end{equation}
In the reservoirs with voltage $V$ with respect to the potential of the
superconductors (which are assumed equal for both superconductors in
order to avoid the ac Josephson effect), $f_L$ obtains the form
\begin{equation}
f_L(E;V,T)=\frac{1}{2}\l[\tanh\l(\frac{E+eV}{2k_B
T}\r)+\tanh\l(\frac{E-eV}{2k_B T}\r)\r].
\label{eq:flres}
\end{equation}
It can be shown\cite{wilhelm,huang} that, in the absence of inelastic
interactions and for energies $E<\Delta$, $f_L$ remains constant
throughout the control wires, and hence the reservoir value can
directly be used for the calculation of the supercurrent.

In this paper, we will consider two limits for $f_L$. These are the
equilibrium finite-temperature limit, where
$f_L=\tanh(E/2k_BT)$, and the zero-temperature nonequilibrium
case when $f_L$ is driven in a normal-metal wire,\cite{wilhelm}
$f_L=\vartheta(E-eV)-\vartheta(-eV-E)$, where 
$\vartheta(E)$ is the Heaviside step function.

The spectrum of supercurrent-carrying states typically consists of
both the states carrying the supercurrent parallel to the phase
gradient and those carrying it in the opposite
direction,\cite{wilhelm,kulik} depending on their energy. Hence, by
controlling the occupation of these states by the above steplike
distribution function, one is able to vary the sign of the observable
supercurrent and, e.g., obtain the $\pi$ state.  

The form of the spectrum can be qualitatively understood by
considering a ballistic (scattering-free) weak link. There, the
quasiparticles form bound
states\cite{kulik,ishii,bardeenjohnson,gunsenheimer,beenakkervanhouten,buttikerklapwijk,vanwees91} 
which contain an Andreev reflection\cite{andreev} at both NS
interfaces. Since the first reflection at the left is from hole- to
particle (particle- to hole) -like states and the second at the right
interface from particle- to hole (from a hole- to particle) -like
states, the net result is a transfer of a Cooper pair from the left
superconductor to the right (from right to left). Bound-state energies
are found by requiring that the total phase the quasiparticles acquire
within a single cycle is a multiple of $2\pi$. This leads to (for $E_m
\ll \Delta$) 
\begin{equation}
E_m^{\pm}=\frac{1}{2\tau}\left[2\pi\left(m+\frac{1}{2}\right)\pm\phi\right],
\label{eq:ballisticboundstates}
\end{equation}
the sign in front of the phase depending on the direction of the
supercurrent flow. Here $\tau=v_F/L$ is the time of flight between two
successive Andreev reflections and $L$ is the corresponding length of
the trajectory. The supercurrent-carrying density of states is then
found from 
\begin{equation}
j_S(E;\phi) \propto \sum_m \frac{\partial E_m^\pm}{\partial
\phi} \delta(E-E_m),
\end{equation}
resulting into a peaklike spectrum that contains states carrying both
positive ($E=E_m^+$) and negative ($E=E_m^-$) supercurrent. In the
presence of disorder, the distribution of the times of flight $\tau$
depends on the impurity potential and the spectral supercurrent is
conveniently characterized by its impurity-averaged smooth density of
states. However, the resulting spectrum still contains many properties
similar to the clean limit, such as the varying sign of the
supercurrent carried at different energies. This analogy holds, even
though on a formal level, the calculation within our quasiclassical
technique does not directly invoke these concepts.

\section{Short- and long-junction limits}
\label{lscales}
The spectrum of current-carrying states in the weak link depends very
much on the ratio of the length $L$ of the weak link and the
superconducting coherence length $\xi_0 = \sqrt{D/2\Delta}$, or in
other words, on the ratio between the superconducting order parameter
$\Delta$ and the Thouless energy $E_T=D/L^2$ of the weak link. 
In the case of a long weak link,\cite{wilhelm} $L \gg \xi_0$ (or,
equivalently, $E_T \ll \Delta$), the spectrum is wide and many energy
states contribute to the supercurrent with only a small
phase-dependent gap of the order of a few $E_T$ at low energies. In
the opposite limit, only the states with energy $E \in [\Delta
|\cos(\phi/2)|,\Delta]$ carry supercurrent and between these limits,
$\Delta$ serves as a cutoff for the spectral supercurrent: there may
exist some current-carrying states with $E>\Delta$, but their
contribution vanishes quickly with $E-\Delta$. 

The energy ranges can be understood as follows: In the center
of the junction, which is the bottleneck for the supercurrent, both 
superconductors provide sufficient correlations that a gap 
in the energy spectrum of a size $E_g$ is induced at $\phi=0$, where
the size of 
$E_g$ interpolates between $E_{\rm T}$ (long junction) and $\Delta$
(short junction).\cite{GolKupr,Belzig} If 
now a finite phase difference is applied, the correlations from either side
start to interfere more and more destructively leading to a closing of the 
gap at $\phi=\pi$.\cite{Charlat}
Hence the lower energy bound, below which no bound states
exist,  is set by this phase-dependent gap. Above $\Delta$, the states 
depend less and less on the superconducting properties, hence their phase
dependence is rapidly lost and they also do not contribute to the 
supercurrent. 

\subsection{Short-junction limit $L \ll \xi_0$}
In the limit when the superconducting order parameter $\Delta$ is much 
smaller than the Thouless energy, the supercurrent is carried by
states with energies much below $E_T$ and we may thus neglect the
first term on the right side of Eq.\ (\ref{eq:usadela}). In this case, 
we get an analytical solution to the differential equations without
further approximations,
\begin{align}
\theta(x)&=\arcosh\left(\frac{\sqrt{\alpha^2+1}}{\alpha}\cosh[j_E \alpha
(x-x_0)]\right),\\ 
\chi(x)&=\chi_0-\arctan\{\alpha \tanh[j_E\alpha(x-x_0)]\},
\end{align}
where $\alpha$ and $x_0$ are constants which along with the spectral
supercurrent $j_E$ are determined from the boundary conditions. In the
two-probe case we can choose the origin in the center of the weak
link, and assume the functions $\theta(x)$ and $\chi(x)$ take the bulk
values at the NS boundary ($x=\pm L/2$). Thus we get $x_0=0$ and 
\begin{align}
\alpha&=\frac{\sqrt{E^2-\Delta^2 \cos^2(\phi/2)}}{\Delta
\cos(\phi/2)},\\
j_E &= \frac{2 \Delta \cos(\phi/2)}{\sqrt{E^2-\Delta^2
\cos^2(\phi/2)}} \arcosh\left(\sqrt{\frac{E^2-\Delta^2
\cos^2(\phi/2)}{E^2-\Delta^2}}\right). 
\end{align}
In the real-time calculation of the supercurrent, we are mostly
interested in the imaginary part of the spectral supercurrent
$j_E$. This is 
\begin{equation}
{\rm Im}(j_E) = 
\begin{cases}
0, \quad E > \Delta\\
\frac{\pi \Delta \cos(\phi/2)}{\sqrt{E^2-\Delta^2\cos^2(\phi/2)}},
\quad E \in [\Delta \abs{\cos(\phi/2)}, \Delta]\\
0, \quad \abs{E} < \Delta \abs{\cos(\phi/2)},
\end{cases}
\label{eq:speccurshort}
\end{equation}
and ${\rm Im}[j_E (-E)]=-{\rm Im}[j_E (E)]$. At $T=0$, we get the
observable supercurrent by simply integrating $\Im(j_E)$ over the
energy to obtain
\begin{equation}
I_S = \frac{\pi \Delta \cos(\phi/2)}{eR_N}
\artanh[\sin(\phi/2)].
\label{eq:shortjunctionT0}
\end{equation}
For a finite temperature, we have to multiply this by the
distribution function $\tanh(E/2k_BT)$ and integrate over the energy,
which is conveniently done using the Matsubara technique [i.e.,
substituting $E=i\omega_n$,
where $\omega_n=\pi T (2n+1)$ are the poles of $\tanh(E/2k_BT)$ and
summing $\Re(j_E)$ over $n=0,1,...$, see Refs.~\onlinecite{Pascal}
and \onlinecite{review} for details], yielding 
\begin{equation}
\begin{split}
I_S =& \frac{2\pi \Delta T}{eR_N} \cos(\phi/2)\sum_{n=0}^\infty
\frac{1}{\sqrt{\Delta^2\cos^2(\phi/2) + \omega_n^2}}\\
&\times {\rm arctan}\l(\frac{\Delta \sin(\phi/2)}{\sqrt{\Delta^2\cos^2(\phi/2)+\omega_n^2}}\r).
\end{split}
\label{eq:shortjunctionT}
\end{equation}
As expected, Eqs.~(\ref{eq:shortjunctionT0}) and (\ref{eq:shortjunctionT})
are the same as obtained by Kulik and
Omel'yanchuk\cite{kulikomelyanchuk} and in numerical
studies\cite{Pascal} for the same limit. 

In a setup where the distribution function can be controlled by an
additional probe coupled to the system via a narrow normal wire (such
that the current-carrying states are not essentially deformed), the
resulting supercurrent as a function of control voltage $V$ at $T=0$
reads 
\begin{equation}
I_S(V)=\frac{\pi\Delta \cos(\phi/2)}{2eR_N}
\ln\left[\frac{\Delta(1+\sin(\phi/2))}{V+\sqrt{V^2-\Delta^2\cos^2(\phi/2)}}\right]
\end{equation}
for $V \in [\Delta \abs{\cos(\phi/2)}, \Delta]$. Above $\Delta$, $I_S$
vanishes, and for $V \le \Delta\abs{\cos(\phi)}$, the supercurrent has
the form of Eq.~(\ref{eq:shortjunctionT0}), independent of $V$.

\begin{figure}[h]
\includegraphics[width=0.9\columnwidth,clip]{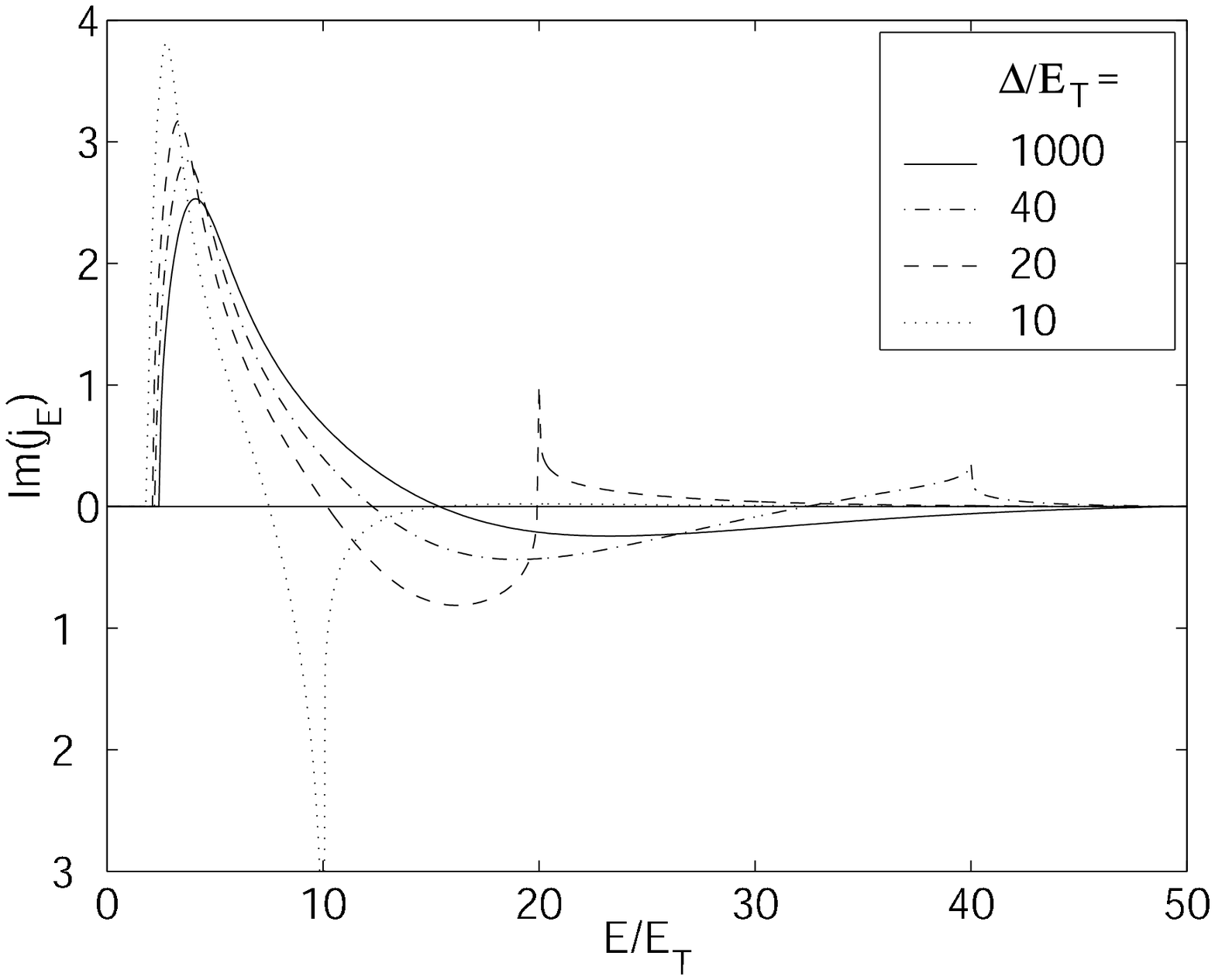}
\includegraphics[width=0.9\columnwidth,clip]{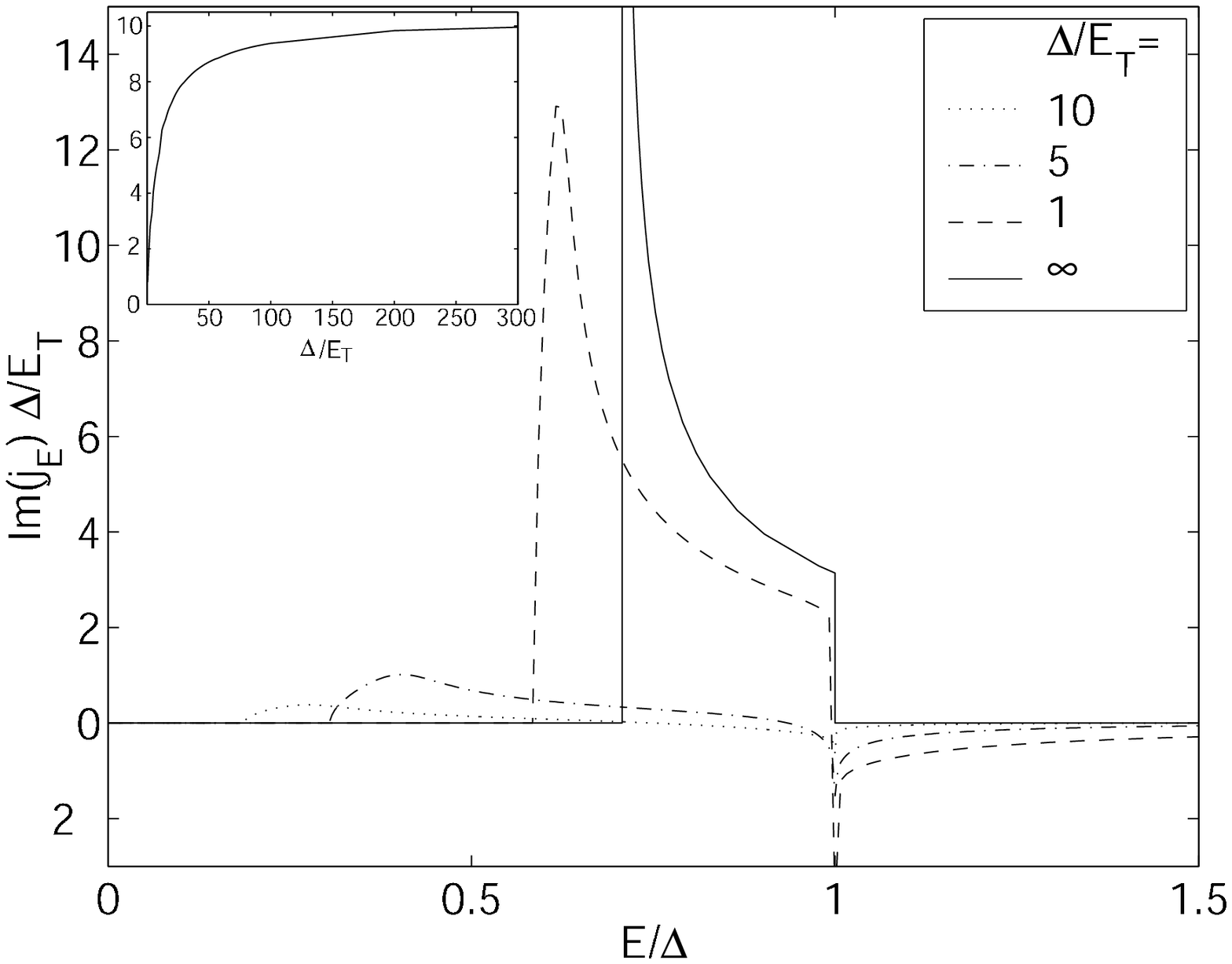}
\caption{Spectral supercurrent for a few values of $\Delta/E_T$. (a)
Junctions longer than the superconducting coherence length $\xi_0$. A
finite $\Delta/E_T$ shows up as a peak at $E=\Delta$. (b) Short
junctions $L \sim \xi_0$: the peak at $E=\Delta$ persists, but another 
develops around $E=\Delta \cos(\phi/2)$. In (b), the spectral
supercurrent is normalized by $E_T/\Delta$ to allow for the analytical 
solution at $E_T/\Delta \rightarrow \infty$. The inset shows how the
zero-temperature, zero-voltage critical current behaves as a function
of $\Delta/E_T$, in accordance with Ref.\ 10.}
\label{fig:speccurdeltadep}
\end{figure}

The spectral supercurrent of Eq.~(\ref{eq:speccurshort}) can also be
obtained from the diffusive limit of the corresponding quantity
derived in Ref.\ \onlinecite{beenakker91}. There, the supercurrent is
written as a sum of the contributions from different bound states,
\begin{equation}
I_S=\frac{e\Delta}{2} \sin(\phi) \sum_{p=1}^{N}
\frac{\tau_p}{E_p} \tanh\left(\frac{E_p}{2k_B T}\right),
\label{eq:beenakker}
\end{equation}
where the bound-state energies $E_p$ depend on the transmission 
eigenvalues $\tau_p$ by $E_p=\Delta [1-\tau_p
\sin^2(\phi/2)]^{1/2}$. Writing Eq.~(\ref{eq:beenakker}) in the
form of an energy integral,
\begin{equation}
I_S=\frac{e\Delta}{2} \sin(\phi) \int dE \sum_{p=1}^{N} 
\frac{\tau_p}{E} \tanh\left(\frac{E}{2k_B T}\right)
\delta(E-E_p),
\label{eq:beenint}
\end{equation}
and averaging the transmission eigenvalues over their diffusive-limit
distribution,\cite{nazarovsl} $\rho(\tau)=(\pi/2e^2 R_N)
(\tau\sqrt{1-\tau})^{-1}$, yields a spectral supercurrent given by
Eq.~(\ref{eq:speccurshort}) multiplying the distribution function
$\tanh(E/2k_BT)$. 

\subsection{Long and intermediate-length junctions}
If the length $L$ of the weak link is much longer than $\xi_0$, the
supercurrent is carried by a wide spectrum of
energies\cite{wilhelm,yip,heikkila}. At low $E$, however, the
current-carrying density of states has a phase-dependent minigap
reminiscent of the gap in the usual density of states of a SNS sample
\cite{gwz}. Above the gap, $\Im (j_E)$ rises sharply, then starts to
oscillate with an exponentially decaying envelope. This oscillatory
behavior is responsible for the occurrence of the $\pi$ state in
nonequilibrium-controlled Josephson junctions\cite{baselmans,huang}
and in ferromagnetic weak links.\cite{buzdin,ryazanov}

The spectral supercurrent as a function of energy is plotted for a few 
values of $\Delta/E_T$ in Fig.\ \ref{fig:speccurdeltadep}, the upper
figure showing the limit $E_T \ll \Delta$ and the lower the limit $E_T 
\lesssim \Delta$.

For a finite ratio $\Delta/E_T$, the divergence of the density of states at 
the superconducting gap edge is reflected as a 
peak in the spectral supercurrent at $E=\Delta$. The direction of the
peak, positive or negative, is determined by geometric considerations
and hence depends on the precise value of $\Delta/E_T$. For $E_T
\approx \Delta$, the spectral supercurrent $\Im(j_E)$ tends towards
the short-junction result, Eq.\ (\ref{eq:speccurshort}), replacing the 
Thouless gap by the gap of width $\Delta \cos(\phi/2)$. Moreover, for
$\Delta/E_T \rightarrow 0$, the width of the peak at $E=\Delta$ tends
to zero.

In the limit $T \gg E_T$ for a long junction ($E_T \ll \Delta$), the
temperature dependence of the obtained observable supercurrent tends
to the limits considered by Likharev\cite{likharev76} for $\Delta \ll
T$ and by Zaikin and Zharkov\cite{zaikin81} for a general $\Delta/T$.

\begin{figure}
\includegraphics[width=0.9\columnwidth,clip]{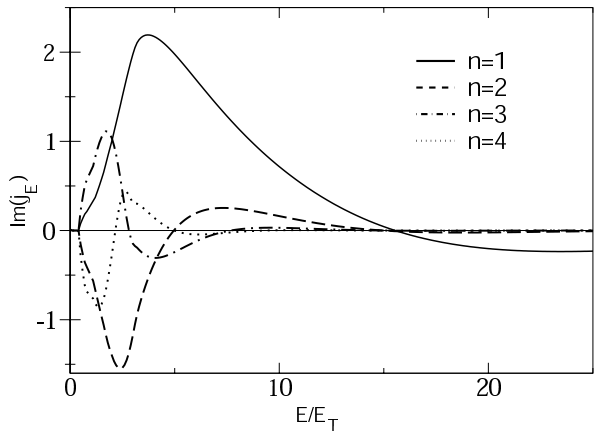}
\caption{Energy dependence of the Fourier sin-transformed spectral
supercurrent: four lowest harmonics $n=1,2,3,4$ corresponding to the
phase dependencies $\sin(n\phi)$. The energy scales and magnitudes of
the different harmonic constituents decay as $1/n$.}
\label{fig:jeharmonics}
\end{figure}

\section{Phase dependence}
\label{phase}

Originally, the Josephson effect was discovered for insulating weak
links in the tunneling regime, i.e., in the limit of
a very low 
tunneling probability. There, the supercurrent is due to an
uncorrelated transfer of Cooper pairs through the weak
link.\cite{FN1} As a result, one obtains the familiar dc Josephson
relation $I_S=I_c \sin(\phi)$. However, it has been shown (see,
e.g., Refs.\ \onlinecite{ishii} and \onlinecite{kulikomelyanchuk})
that other kinds of weak links, through which the transmission
probability is much above zero, may have a different current-phase
relation. Thus we may write in general 
\begin{equation}
I_C(\phi) = \sum_{n=1}^\infty I_C^n \sin(n\phi),
\label{eq:harmonics}
\end{equation}
where the amplitudes $I_C^n$ are the coefficients of the Fourier
sine series of $I_C(\phi)$. For example, in a ballistic weak link where 
the transmission probability for Cooper pairs is 1, $I_C^n \propto
-(-1)^n/n$, yielding a sawtooth form for
$I_C(\phi)$.\cite{ishii,vanwees91} The odd parity with respect to
$\phi$ 
(appearance of only sine terms) of this representation reflects the
fact that the supercurrent is driven by the spatial asymmetry 
introduced by the application of the phase: changing the phase to a negative
value corresponds to mirroring the structure about the center and 
hence to a reversal of the current. 

The occurrence of higher harmonics in Eq.\ (\ref{eq:harmonics}) may be
interpreted as a correlated transfer of $n$ Cooper pairs through the
weak link as a result of the pairing correlations extending through
quasiparticle paths containing multiple Andreev reflections. For
example, the flux quantum for the $n$th harmonic is $h/2ne$, i.e.,
corresponding to a charge $2ne$.

\begin{figure}
\includegraphics[width=0.9\columnwidth,clip]{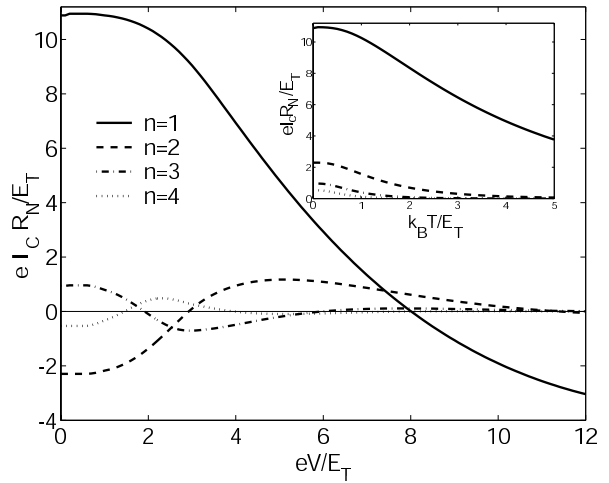}
\caption{Voltage dependence of the amplitudes of the first four 
harmonics of the observable supercurrent. Inset: Corresponding
temperature dependence (even harmonics yield negative amplitudes, but
here we plot the absolute values of the supercurrent).}
\label{fig:Isharm}
\end{figure}

In a diffusive weak link considered in this paper, the transmission
probabilities for the Cooper pairs are widely distributed 
between zero and 1.\cite{nazarovsl} As a
result, one may get contributions from the higher harmonics to the
phase dependence. This is shown in Fig.\ \ref{fig:jeharmonics} where
the amplitudes $\Im(j_E^n)$ of the first four harmonics of the Fourier
sine transformed spectral supercurrent through a long weak link ($L \gg 
\xi_0$) are plotted as a function of energy. Both the amplitude of
$\Im(j_E^n)$ and the effective energy scales decay with a power of $n$
suggesting that the observation of the higher harmonics is easiest at
low temperatures. The corresponding temperature and voltage
dependencies of the critical currents $I_C^n$ plotted in Fig.\
\ref{fig:Isharm} behave analogously. A numerical fit to the obtained
critical currents at $eV=k_BT=0$ yields roughly $I_C^n \propto
-(-1)^n/n^2$ and to the voltage $V^*_n$ where $I_C^n(V)$ first changes
sign suggests that the effective energy scales behave as $E_n^*=E_T(c_1 +
c_2/n)$, with some constants $c_{1,2}$.

This behavior can be understood by identifying the higher harmonics with
the correlated transfer of a cluster of $n$ Cooper pairs. Now instead
of the phase $\phi$, the cluster has the phase $n\phi$ and since the
cycle contains $2n$ Andreev reflections, the effective trajectory
length is increased from $L$ to $nL$. In
Eq.~(\ref{eq:ballisticboundstates}), making these replacements yields
the observed result, $E_n^* \propto (c_1+c_2/n)$. In the diffusive
limit, the effective trajectory length increases in the second power
of the length of the weak link,  $l_{\rm eff} \propto L^2$, but since
the phase is reset after each traversal through the weak link, we
simply get $l_{\rm eff,n} \propto nL^2$. Hence, similarly to the
alternating sign, the scaling of the effective energies with index $n$ 
follows the behavior of the ballistic-limit spectral supercurrent.



Since the crossover voltages $V^*_n$, where $I_C^n(V^*_n)=0$, depend
on $n$, the actual critical current never vanishes at the crossover:
it is rather that the current-phase relation changes its form near the 
crossover voltages. Such a change was observed in Ref.\
\onlinecite{baselmansphase}, where the current-phase relation of a 
controllable Josephson junction was measured in a superconducting
quantum interference device (SQUID) geometry. 

In the short-junction regime $L \ll \xi_0$ the 
contributions of the different harmonics can be derived 
analytically. A general form
for $\Im (j_E^n)$ would be complicated, but as an example, the first two 
amplitudes are
\begin{align}
\Im(j_E^1)&=\left(\frac{E}{\Delta}\right)^2,\\
\Im(j_E^2)&=-\frac{E^2 (2\Delta^2 - 3E^2)}{\Delta^4}.
\end{align}
Clearly all the harmonics share the same energy scale, $\Delta$,
but as for the case of a long weak link, the amplitude decays also
here roughly as $1/n^2$. Namely, integrating $\Im(j_E^n)$ over the
energy, we get for a short weak link 
\begin{equation}
I_C^n = -\frac{(-1)^n e\Delta}{R_N(2n+1)(2n-1)}.
\end{equation}
Replacing $\Delta$ by $E_T$ and scaling by a numerical factor close to
33, this form fits the amplitudes of the long-junction harmonics as
well but a rigorous proof does not exist.

\begin{figure}
\includegraphics[width=0.9\columnwidth,clip]{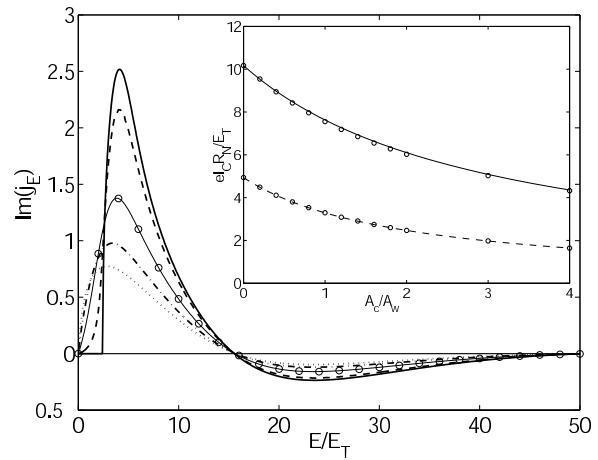}
\caption{Spectral supercurrent for different cross sections $A_c$ of the
control probe. From top to bottom, the cross section is
$A_c=0,0.2,1,2,4$ times the cross section of the weak link. Inset:
observable supercurrent as a function of $A_c$ for $T=V=0$ (upper
set of circles) and for $k_BT=3E_T$ (lower set of circles). The
zero-temperature result is fitted to $I_S(A_c=0)A_w/(A_w+A_c/3)$
(solid line) and the finite-temperature result to
$I_S(A_c=0)A_w/(A_w+A_c/2)$ (dashed line).}
\label{fig:speccurvswidth}
\end{figure}

\section{Extra terminals}
\label{extraterminals}
In order to relate our results to physical observables, we have to
evaluate statistical expectation values. In two-probe SNS weak links
in equilibrium, most of the experimental observations have 
been accurately described with the equilibrium Matsubara
technique.\cite{Pascal} However, one of the recent advances in the
research of the Josephson effect has  
been done in nonequilibrium situations where the distribution function 
in the weak link has been controlled by coupling one or more
normal-metal reservoirs to the weak link by phase-coherent
wires.\cite{baselmans,huang} While making it possible to control the 
occupation of the current-carrying states, these extra wires also
affect the form of $\Im(j_E)$.\cite{yip} In the discussion of these
effects, we concentrate on the regime of a long junction, $L \gg
\xi_0$. There, 
most notably, the control probes allow for the existence of states
with low energies, and therefore the Thouless gap is
lifted. Moreover, the existence of the normal reservoirs brings some
extra depairing by imposing a vanishing boundary condition for the
pairing amplitude $f=\sinh(\theta)$ inside the reservoirs. As a
result, the amplitude of the spectral supercurrent decreases due to
the extra probes. In what follows, we consider the effects of the
integrated cross-sectional area $A_c$ of the control wires attached to 
the center of the weak link with cross section $A_w$ (note: a similar
effect would be present if the control wires and the weak link were
made of different materials with normal-state conductivities
$\sigma_{N,c}$ and $\sigma_{N,w}$ --- however, here we simply
talk about $A_c$ and $A_w$) and of the length $L_c$ of the phase
coherent control wires, compared to the length $L$ of the weak
link. For simplicity, we assume that the widths of the control wires are
much smaller than $L$, allowing one to treat these wires as quasi-1D
structures, connected by the rules of Nazarov's circuit
theory\cite{nazarovsl,nazarovbc}. In the language of this circuit
theory, the extra normal wires divert some of the spectral current to
the normal reservoirs, thus decreasing the pairing correlations
between the superconductors.

\begin{figure}
\includegraphics[width=0.9\columnwidth,clip]{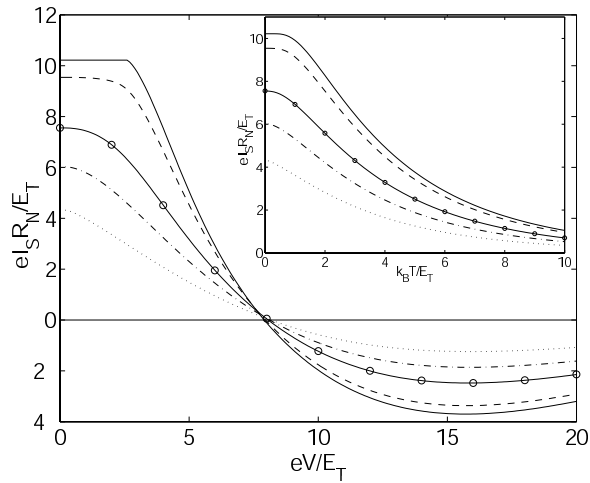}
\caption{Voltage dependence of supercurrent for $\phi=\pi/2$ in
the presence of the control probe with different cross sections
$A_c$. Inset: The corresponding temperature dependence of the
supercurrent. The cross sections have been chosen as in
Fig. \ref{fig:speccurvswidth}.}
\label{fig:Isvswidth}
\end{figure}

The spectral supercurrent as a function of energy for different cross
sections $A_c$ of the control wire is plotted in
Fig.\ \ref{fig:speccurvswidth}. Here we have taken the length $L_c=5L 
\gg L$. Already a small $A_c \ll A_w$ yields a finite $\Im(j_E)$ at
low energies, but does not much reduce the total magnitude of the
supercurrent. The resulting temperature and voltage dependencies of
the total supercurrent $I_s$ is plotted in Fig.\
\ref{fig:Isvswidth}. Except at the very lowest voltages/temperatures
of the order of the Thouless gap, the extra probes do not change the
voltage/temperature dependence of $I_s$ from the two-probe case, but
only the overall magnitude is decreased. From the resulting
$I_S(A_c)$ we obtain 
\begin{equation}
I_S(A_w,A_c) = \frac{A_w}{A_w+A_c/2} I_S(A_w,A_c=0) ,
\end{equation}
which holds very well for $\max(eV, k_B T) \gg E_T$.

\begin{figure}
\includegraphics[width=0.9\columnwidth,clip]{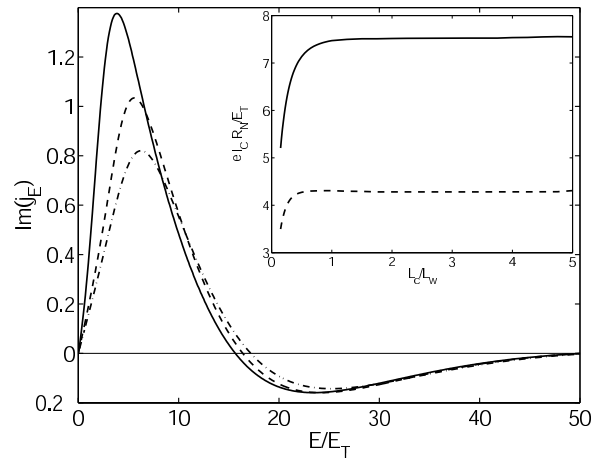}
\caption{Spectral supercurrent for different lengths $L_c$ of the
control probe. From top to bottom, the length is $L_c=5, 0.25, 0.15$
times the length of the weak link. The cross section of the control
probe is chosen equal to that of the weak link. Inset: Supercurrent at
$T=0$ (solid line) and $T=3E_T/k_B$ (dashed) as a function of the
length $L_c$ of the control probe with respect to the length of the
weak link.} 
\label{fig:speccurvslength}
\end{figure}

\begin{figure}
\includegraphics[width=0.9\columnwidth,clip]{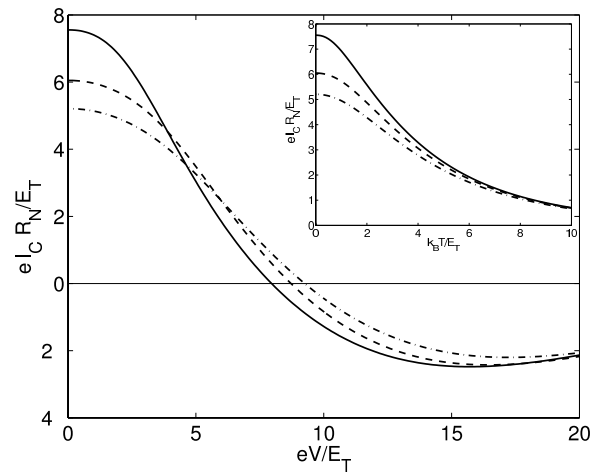}
\caption{Voltage dependence of supercurrent for $\phi=\pi/2$ in 
the presence of a control probe with length $L_c$ and cross section
equal to that of the weak link. Inset: The corresponding temperature
dependence of the supercurrent. The lengths $L_c$ have been chosen as
in Fig. \ref{fig:speccurvslength}.}
\label{fig:Isvslength}
\end{figure}

If the length $L_c$ of the phase-coherent control wire is larger than
$L$, the effect of the control wires is independent of the precise
value of $L_c$. For $L_c < L$, the spectral supercurrent is altered
for $E \lesssim \hbar D/L_c^2$ such that the overall magnitude is
decreased, the observable supercurrent tending towards zero as 
$L_c \rightarrow 0$. The spectral supercurrent and the resulting
temperature/voltage dependencies for $A_c=A_w$ and for three different
$L_c/L$ are plotted in Figs.\ \ref{fig:speccurvslength} and
\ref{fig:Isvslength}, respectively. 

\begin{figure}
\includegraphics[width=0.88\columnwidth,clip]{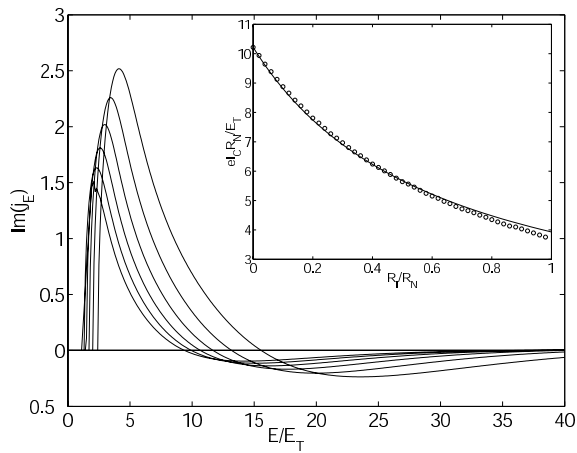}
\caption{Spectral supercurrent in a weak link with dirty interfaces
to the superconductors, characterized by the ratio $r_b=R_I/R_N$ of the
resistances. The interface resistance $R_I$ is assumed the same for
both interfaces. From top to bottom: $r_b=0$, 0.2, 0.4, 0.6, 0.8,
1.0. Enhanced scattering at the interface reduces mostly the
amplitude of the supercurrent, but also slightly the effective energy
scale. Inset: zero-temperature, zero-voltage supercurrent as a
function of $r_b$. Circles: calculated supercurrent;  solid line: fit
to $I_S(r_b) = I_S(r_b=0)/(1+1.6r_b)$.} 
\label{fig:jevsrbdirty}
\end{figure}

\begin{figure}
\includegraphics[width=0.88\columnwidth]{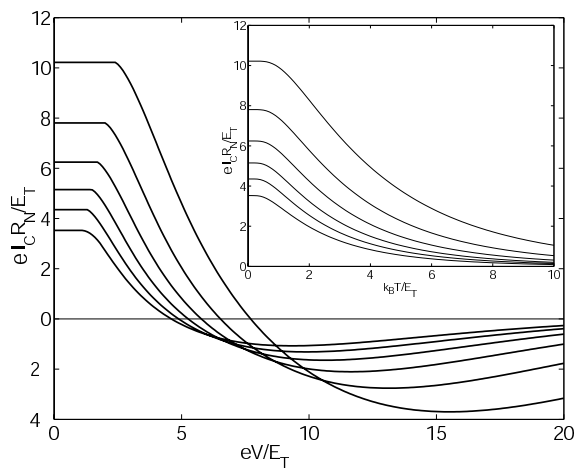}
\caption{Voltage dependence of the supercurrent with dirty
NS interfaces at $\phi=\pi/2$. Inset: Corresponding temperature
dependence. The values of $R_I$ are the same as in Fig.\
\ref{fig:jevsrbdirty}.} 
\label{fig:Isvsrbdirty}
\end{figure}

\section{Nonideal interfaces}
\label{interfaces}

Normal-metal --- superconductor interfaces with reduced transmissivity
can be taken into account by specifying the transmission eigenvalues
through the interface and taking them into account as in
Eq.~(\ref{eq:nazarovbcforgf}). Here we consider a typical case
described by the distribution of eigenvalues for a ``dirty''
interface. 

Figure \ref{fig:jevsrbdirty} shows the spectral supercurrent for a
long junction connected to superconductors through a dirty interface
with resistance $R_I$ (yielding a total resistance $R_N+2R_I$ between
the superconductors). Due to the additional resistance, the
amplitude of the supercurrent decays with $r_b$, but also the
energy scales decrease since the interface barrier can to certain
extent be thought as adding a barrier-equivalent length\cite{zhou95}
to the path length of the quasiparticles. Observing the 
temperature and voltage dependencies of the resulting supercurrent,
plotted in Fig.~\ref{fig:Isvsrbdirty}, shows that the amplitude of the 
supercurrent behaves at $T=0$, $V=0$ as
\begin{equation}
I_C(R_I,R_N) \approx \frac{I_C(R_I=0)R_N}{R_N+1.6 R_I},
\end{equation}
i.e., the resistances should not simply be added up but the dirty
interface decreases the supercurrent less efficiently than the
normal-metal resistance. Furthermore, the effective energy scale $E^*$
found, e.g., from the voltage dependence indicates that it follows the
approximate law $E^*=E_T(1+0.7R_I/R_N)$. This kind of a behavior of
the critical current and the spectral supercurrent is similar to those
found in Ref.~\onlinecite{kuprianovlukichev} [especially, see
Eqs.~(55) and (56)] and Ref.~\onlinecite{yip} in the case of a
tunneling interface. 

\section{Conclusions}
\label{summary}
In this paper, we have systematically investigated the spectrum of
current-carrying states in a phase coherent normal-metal weak
link. Taking into account the effect of extra terminals, the
characteristic energy scales --- the Thouless energy $E_T$ and the
BCS superconducting gap $\Delta$ --- and a finite NS interface
resistance makes it easier to find a quantitative agreement with
the obtained experimental results on the nonequilibrium-controlled
supercurrent. We have also been able to derive analytical results in 
a number of limits. Moreover, we have discussed the underlying 
microscopic phenomena leading to the $\pi$ state and have 
explained its properties, such as its dependence on energy, and higher 
harmonics in the phase dependence, by invoking Andreev bound states
smeared over a broad distribution of times of flight, and by multiple
Andreev cycles tranferring more than one Cooper pair in a single
coherent process.

To obtain an optimal voltage control of the supercurrent, the
interface resistances should be much smaller than the weak-link 
wire resistance, the control wire should be slightly longer than the
weak link (but not much longer to reduce the inelastic effects on the
nonequilibrium distribution function), and as thin as possible
compared to the weak link. For the observation of the $\pi$ state, the
ideal limit is the long-junction limit $L \gg \xi_0$, where the
spectrum of the current-carrying states is not cut off by the
superconducting gap.  

In typical experiments, also inelastic scattering neglected here may
become important. However, since equilibrium phenomena induced by the
superconducting proximity effect have been quantitatively described by
the quasiclassical theory without incorporating such effects (see,
e.g., Refs.\ \onlinecite{Pascal} and \onlinecite{sillanpaa}), we expect these inelastic
effects to be mostly important in the kinetic equations describing the 
nonequilibrium distribution functions. In recent
experiments,\cite{baselmanssquid,huang} including these inelastic
terms into the kinetic equations has lead to good agreement between
the theory and the experiments. Therefore our results provide an
accurate and independent way of also characterizing such inelastic
effects by observing how they affect the supercurrent.

\section*{ACKNOWLEDGEMENTS}
We thank Norman Birge, Fr$\acute{\rm e}$deric Pierre and
Jochem Baselmans for discussions. This work was supported by the
Graduate School in Technical Physics at the Helsinki University of
Technology.

\bibliography{basename of .bib file}

\begin{thebibliography}{99}
\bibitem{wilhelm} F. K. Wilhelm, G. Sch\"on, and A. D. Zaikin,
Phys.\ Rev.\ Lett.\ {\bf 81}, 1682 (1998).
\bibitem{yip} S. K.\ Yip, Phys.\ Rev.\ B {\bf 58}, 5803 (1998).
\bibitem{volkov} A. F.\ Volkov, Phys.\ Rev.\ Lett. {\bf 74}, 4730 (1995);
A. F.\ Volkov, R.\ Seviour, and V. V.\ Pavlovskii, Superlattices Microstruct.\ 
{\bf 25}, 647 (1999). 
\bibitem{morpurgo} A. F.\ Morpurgo, T. M. Klapwijk, and B. J.\ van Wees,
Appl.\ Phys.\ Lett. {\bf 72}, 966 (1998).
\bibitem{baselmans} J. J. A.\ Baselmans, A. F.\ Morpurgo, B. J. van Wees, 
and T. M.\ Klapwijk, Nature (London) {\bf 397}, 43 (1999); J. J. A. Baselmans,
B. J. van Wees, and T. M. Klapwijk, Phys.\ Rev.\ B {\bf 63}, 094504 (2001). 
\bibitem{huang} J.\ Huang, F.\ Pierre, T. T.\ Heikkil\"a, F. K.\ Wilhelm,
and N. O.\ Birge, Phys. Rev. B {\bf 66}, 020507(R) (2002). 
\bibitem{schaepers} Th.\ Sch\"apers {\em et al.}, Appl.\ Phys.\ Lett.\ 
{\bf 73}, 2348 (1998).
\bibitem{baselmanssquid} J. J. A. Baselmans, B. J. van Wees, and
T. M. Klapwijk, Phys. Rev. B {\bf 65}, 224513 (2002) .
\bibitem{kutchinsky} J. Kutchinsky, R. Taboryski, C. B. S\o rensen,
J. B. Hansen, and P. E. Lindelof, Phys. Rev. Lett. {\bf 83}, 4856
(1999). 
\bibitem{Pascal} P.\ Dubos, H.\ Courtois, B.\ Pannetier, F. K.\ Wilhelm,
A. D.\ Zaikin, and G.\ Sch\"on, Phys.\ Rev.\ B {\bf 63}, 064502
(2001). 
\bibitem{baselmansphase} J. J. A. Baselmans, T. T. Heikkil\"a,
B. J. van Wees, and T. M. Klapwijk, Phys. Rev. Lett. {\bf 89}, 207002 (2002).
\bibitem{kulik} I. O. Kulik, Zh. Eksp. Teor. Fiz. {\bf 57}, 1745
(1969) [Sov. Phys. JETP {\bf 30}, 944 (1970)].  
\bibitem{ishii} Ch. Ishii, Prog. Theor. Phys. {\bf 44}, 1525 (1970).
\bibitem{bardeenjohnson} J. Bardeen and J. L. Johnson, Phys. Rev. B
{\bf 5}, 72 (1972).
\bibitem{buttikerklapwijk} M. B\"uttiker and T. M. Klapwijk,
Phys. Rev. B {\bf 33}, 5114 (1986).
\bibitem{vanwees91} B. J. van Wees, K.-M. H. Lenssen, and
C. J. P. M. Harmans, Phys. Rev. B {\bf 44}, 470 (1991).
\bibitem{beenakkervanhouten} C. W. J. Beenakker and H. van Houten,
Phys. Rev. Lett. {\bf 66}, 3056 (1991).
\bibitem{gunsenheimer} U. Gunsenheimer, U. Sch\"ussler, and
R. K\"ummel, Phys. Rev. B {\bf 49}, 6111 (1994).
\bibitem{usadel} K. D.\ Usadel, Phys.\ Rev.\ Lett.\ {\bf 25}, 507
(1970). 
\bibitem{review} For an introduction see, e.g., W.\ Belzig, F. K.\ Wilhelm, 
C.\ Bruder, G.\ Sch\"on, and A. D.\ Zaikin, Suplatt. Microstruct. 
{\bf 25}, 1251 (1999). 
\bibitem{rammersmith} J. Rammer and H. Smith, Rev. Mod. Phys. {\bf 58}, 323
(1986).
\bibitem{solsferrer} F. Sols and J. Ferrer, Phys. Rev. B {\bf 49},
15 913 (1994).
\bibitem{zaitsev} A. V.\ Zaitsev, Zh.\ Eksp.\ Teor.\ Phys.\ {\bf 86},
1742 (1984) [Sov.\ Phys.\ JETP {\bf 59}, 1015 (1984)].
\bibitem{kuprianovlukichev} M.\ Yu.\ Kuprianov and V. F.\ Lukichev, 
Zh.\ Eksp.\ Teor.\ Fiz.\ {\bf 94}, 139 (1988) [Sov.\ Phys.\ JETP 
{\bf 67}, 1163 (1988)].
\bibitem{nazarovbc} Yu.\ V.\ Nazarov, Superlattices Microstruct.\ {\bf 25}, 
1221 (1999). 
\bibitem{belzigspin} W. Belzig, A. Brataas, Yu. V. Nazarov, and
G. E. W. Bauer, Phys. Rev. B {\bf 62}, 9726 (2000).
\bibitem{schep} K. M. Schep and G. E. W. Bauer, Phys. Rev. Lett. {\bf
78}, 3015 (1997). 
\bibitem{zaitsevmc} A. V.\ Zaitsev, Physica B {\bf 203}, 274 (1994).
\bibitem{gwz} A. A.\ Golubov, F. K.\ Wilhelm, and A. D.\ Zaikin, Phys.\ 
Rev.\ B {\bf 55}, 1123 (1997). 
\bibitem{andreev} A. F. Andreev, Zh. Eksp. Teor. Fiz. {\bf 46}, 1823
(1964) [Sov. Phys. JETP {\bf 19}, 1228 (1964)]; {\bf 49}, 655 (1965) 
[{\bf 22}, 455 (1966)].
\bibitem{GolKupr} A. A.\ Golubov and M.\ Yu.\ Kupriyanov, J.\ Low Temp.\
Phys.\ {\bf 70}, 83 (1988). 
\bibitem{Belzig} W.\ Belzig, C.\ Bruder, and G.\ Sch\"on, Phys.\ Rev.\ 
B {\bf 54}, 9443 (1996).
\bibitem{Charlat} F.\ Zhou, P.\ Charlat, and B.\ Pannetier, J.\ Low 
Temp. Phys.\ {\bf 110}, 841 (1998). 
\bibitem{kulikomelyanchuk} I. O.\ Kulik and A. N.\ Omelyan'chuk, Fiz. Nizk. Temp. {\bf 4}, 296 (1978) [Sov.\ J.\ Low Temp.\ Phys.\ {\bf 4}, 142 (1978)]. 
\bibitem{beenakker91} C. W. J. Beenakker, Phys. Rev. Lett. {\bf 67},
3836 (1991); {\bf 68}, 1442(E) (1992).
\bibitem{nazarovsl} Yu.\ V.\ Nazarov, Phys.\ Rev.\ Lett.\ {\bf 73}, 134 
(1994). 
\bibitem{heikkila} T. T. Heikkil\"a, F. K. Wilhelm, and G. Sch\"on,
     Europhys. Lett. {\bf 51}, 434 (2000). 
\bibitem{buzdin} A. I. Buzdin, L. N. Bulaevskii, and V. Panyukov,
Pis'ma Zh. Eksp. Teor. Fiz. {\bf 35}, 147 (1982) [JETP Lett., {\bf
35}, 178 (1982)]; A. I. Buzdin and M. Yu. Kupriyanov, {\it ibid.} {\bf 
52}, 1089 (1990) [{\bf 52}, 487 (1990)]; {\bf 53}, 308 (1991) [{\bf
53}, 321 (1991)]; A. I. Buzdin, B. Bujicic, and M. Yu. Kupriyanov,
Zh. Eksp. Teor. Fiz. {\bf 101}, 231 (1992) [Sov. Phys. JETP, {\bf 74},
124 (1992)].  
\bibitem{ryazanov}
V. V. Ryazanov, V. A. Oboznov, A. Yu. Rusanov,  
A. V. Veretennikov, A. A. Golubov, and J. Aarts, Phys. Rev. Lett. {\bf
86}, 2427 (2001). 
\bibitem{likharev76} K. K. Likharev, Pis'ma Zh. Tekh. Fiz. {\bf 2}, 29 
(1976) [Sov. Tech. Phys. Lett. {\bf 2}, 12 (1976)].
\bibitem{zaikin81} A. D. Zaikin and G. F. Zharkov,
Fiz. Nizk. Temp. {\bf 7}, 375 (1981) [Sov. J. Low
Temp. Phys. {\bf 7}, 184 (1981)].
\bibitem{FN1} ``Uncorrelated'' means in this context that no microscopic
correlation between the different Cooper pairs crossing the junction exist. 
The transport is, however, still coherent in the sense that each Cooper pair
keeps its internal correlation while traveling the junction. Furthermore,
the phases of the reservoirs are assumed not to fluctuate such that the
Cooper pairs are transported with the same macroscopic phase and hence 
add up to a macroscopic steady supercurrent.
\bibitem{zhou95} F. Zhou, B. Spivak, and A. Zyuzin, Phys. Rev. B {\bf
52}, 4467 (1995).
\bibitem{sillanpaa} M. A. Sillanp\"a\"a, T. T. Heikkil\"a,
R. K. Lindell, and P. J. Hakonen, Europhys. Lett. {\bf 56}, 590 (2001).
\end{thebibliography}

\end{document}